\documentclass[a4paper,11pt]{article}
\pdfoutput=1 

\usepackage{jcappub} 

\usepackage[T1]{fontenc} 

\makeatletter
\def\@fpheader{\relax}
\makeatother

\title{\boldmath Non-minimally coupled quartic inflation with Coleman-Weinberg one-loop corrections in the Palatini formulation}

\author[a,b]{Nilay Bostan}

\affiliation[a]{Department of Physics, Mimar Sinan Fine Arts University, \\34380 \c{S}i\c{s}li, \.{I}stanbul, Turkey}
\affiliation[b]{Department of Physics and Astronomy, University of Iowa, \\52242 Iowa City, IA, USA }

\emailAdd{nilay-bostan@uiowa.edu}

\abstract{We discuss how the non-minimal coupling $\xi \phi^2 R$ between the inflaton and the Ricci scalar affects predictions of single field inflation models in Palatini formalism. To transition radiation dominated era, the inflaton field $\phi$ must interact to matter fields at the end of inflation. Interactions of the inflaton with other fields lead to radiative corrections to the inflationary potential. These radiative corrections can be explained at leading order by Coleman-Weinberg (CW) one-loop corrections. In this work, using two different prescriptions debated in the literature, the effect of radiative corrections to the potential owing to the coupling of the inflaton to bosons in Prescription I and couplings of the inflaton to bosons and fermions in Prescription II have been examined. We analyze the range of these coupling parameter values for which the spectral index $n_s$ and the tensor-to-scalar ratio $r$ are compatible with the data taken into account to the Keck Array/BICEP2 and Planck collaborations. Finally, we also show that for all the considered potentials the running of the spectral index $\alpha=\mathrm{d} n_s/\mathrm{d} \ln k$ as a function of $\kappa$ for selected $\xi$ values.}

\keywords{Palatini, loop corrections, non-minimal coupling}

\begin{document}
\maketitle
\flushbottom

\section{Introduction}
\label{sec:intro}
Inflation \cite{Guth:1980zm,Linde:1981mu,Albrecht:1982wi,Linde:1983gd}, which is an accelerated expansion era considered to occur in the early Universe, has been widely accepted over the past few decades. It is thought that inflation has become a solution to the various problems of early Universe such as horizon, flatness and monopole. The simplest realization of an inflationary scenario is based on slow-rolling scalar field $\phi$ which is known as inflaton, over a flat potential $V(\phi)$. A large number of inflationary models have been proposed and most of them determining by the inflaton so far. These models predictions are being tested by the cosmic microwave background radiation temperature anisotropies and polarization observations that have become even more precisely in recent years  \cite{Aghanim:2018eyx,Akrami:2018odb}. The latest
data from the Keck Array/BICEP2 and Planck collaborations \cite{Ade:2018gkx} constraints  robustly the tensor-to-scalar ratio $r$, which gives plausible explanation to the amplitude of primordial gravitational waves and the scale of inflation. 

The observational parameters, especially the scalar spectral index $n_s$ and the tensor-to-scalar ratio $r$, have been calculated for various inflationary potentials \cite{Martin:2013tda}. Generally, assuming that the inflaton is coupled to gravitation just through the metric. On the one hand, the action in general also comprises of $\xi \phi^2 R$ coupling term between the Ricci scalar and the inflaton, it is necessary to provide the renormalizability of the scalar field theory in curved space-time \cite{Callan:1970ze,Freedman:1974ze,Buchbinder:1992rb}, and inflationary predictions are substantially changed depending on the coefficient of this coupling term \cite{Abbott:1981rg,Spokoiny:1984bd,Lucchin:1985ip,Futamase:1987ua,Fakir:1990eg,Salopek:1988qh,Amendola:1990nn,Faraoni:1996rf,Faraoni:2004pi}. Here we will investigate how the value of the non-minimal coupling parameter $\xi$ affects the $n_s$, $r$ and $\mathrm{d} n_s/\mathrm{d} \ln k$ for radiatively corrected quartic inflation with non-minimal coupling in Palatini formulation for prescription I and prescription II. In literature, a vast majority of articles take into account to the inflation with non-minimal coupling in Metric formalism \cite{Bezrukov:2010jz,Bostan:2018evz,Bezrukov:2007ep}. However, in this work, we will discuss inflation with non-minimal coupling in Palatini formalism. Unlike Metric formalism, in Palatini, the inflaton remains sub-Planckian regime thanks to provide a natural inflationary epoch \cite{Bauer:2008zj}. Furthermore, for inflation with a non-minimally coupled scalar field, the Palatini formulation leads to different predictions for cosmological parameters \cite{Bauer:2008zj} in particular, the attractor behaviour leading to the predictions of the Starobinsky model is disappear, and r can be much smaller compared to the Metric formulation \cite{Bauer:2008zj,Jarv:2017azx}. In the metric formulation of General Relativity \cite{Padmanabhan:2004fq,Paranjape:2006ca}, the metric and its first derivatives are independent variables, whereas in the Palatini formulation \cite{Attilio, Einstein, Ferraris}, the metric and the connection are the independent variables. Even though the two formalism has the same EoM and thus they correspond to the equivalent physical theories, presence of the non-minimal coupling between gravity and matter, physical equivalence is lost. Therefore, Metric and Palatini formulations describe two different theories of gravity, such refs. recently investigated 
\cite{Bauer:2008zj,York:1972sj,Tenkanen:2017jih,Rasanen:2017ivk,Racioppi:2017spw,Tamanini:2010uq}. 

By taking into account the Palatini formulation is not an additional assumption about the theory, only a different parametrisation of the gravitational
degrees of freedom. It could be discussed that the Palatini formulation is easier than the Metric formulation, since the action does not include any boundary term, as it comprises only first derivatives of the variables. In particular, the attractor behavior of namely as $\xi$-attractor models is vanished in the Palatini formulation \cite{Kallosh:2013tua}. In addition to this, it has been showed that quantum corrections to inflationary potential may play such an important role \cite{Marzola:2015xbh,Marzola:2016xgb,Dimopoulos:2017xox}, if the existence of non-minimal coupling to gravity, leading to the linear inflation \cite{Racioppi:2017spw,Barrie:2016rnv,Kannike:2015kda,Artymowski:2016dlz} and producing the Planck scale dynamically \cite{Kannike:2015kda}.
In literature, inflation with non-minimal coupling in Palatini formalism has been discussed in refs. \cite{Bauer:2008zj,Rasanen:2017ivk,Racioppi:2017spw,Fu:2017iqg,Gumjudpai:2016ioy,Markkanen:2017tun}. Self-interaction potential $V(\phi)$ in Metric and Palatini formulation analyzed in ref. \cite{Bauer:2008zj} and they obtained $n_s\simeq0.968$ and $r\simeq 10^{-14}$ in the large field limit for Palatini approach. In ref. \cite{Rasanen:2017ivk} considered the Higgs inflation in Palatini formulation and they found tensor-to-scalar ratio spanning the range $1\times10^{-13}<r<2\times 10^{-5}$, so $r$ is highly suppressed in Palatini approach. According to the \cite{Racioppi:2017spw}, when $\xi$ increases, $r$ declines and saturating the linear limit for $\xi\gtrsim 10^{-1}$ to the Coleman-Weinberg (CW) inflation in Palatini formalism and also for $\xi\gtrsim1$ values Metric and Palatini formalism discriminate from each other as well as that the Palatini formulation foresees a smaller (larger) value for $r$ $(n_s)$ than the Metric one. Moreover, for $\xi\simeq1$, Palatini approach  gives $r\simeq 0.075$ to the CW inflation with regard to ref. \cite{Racioppi:2017spw}. 

In this paper aims to extend the previous work of non-minimal coupling in Palatini formulation, presenting a non-minimally coupled radiatively corrected quartic inflation potential in Palatini approach. We take into account to the effect of radiative corrections for general values of $\xi\lesssim10^3$, including the case of $\xi\ll1$ by considering both prescription I for bosons coupling and prescription II for bosons and fermions coupling. In ref. \cite{Okada:2010jf}, the inflaton is supposed to couple to fermions and prescription II is used in Metric formulation includes the cases of $\xi\ll1$ and $\xi\simeq2\times10^2$.  In contrast, \cite{Racioppi:2018zoy} considered a potential which encounters with the potential for inflaton to bosons coupling in prescription II for the Metric formulation and further ref. \cite{Bostan:2019fvk} indicated for two different prescriptions to the inflaton couplings to bosons and fermions in Metric formulation. Different from these works, we illustrate the observational parameter values of the regions in the $\kappa-\xi$ plane for which the spectral index and tensor-to-scalar ratio $r$ values are compatible with the current observations and also the running of the spectral index $\mathrm{d} n_s/\mathrm{d} \ln k$ as a function of $\kappa$ for selected $\xi$ values for radiatively corrected quartic inflation potential in Palatini formalism.

To summarize this paper, we first explain the inflation with non-minimal coupling and the definiton of observational parameters which are namely as the spectral index $n_s$, the tensor-to-scalar ratio $r$ and the running of the spectral index $\mathrm{d} n_s/\mathrm{d} \ln k$ in the Palatini formalism (section \ref{inf}). Next, we introduce radiative corrections: prescription I and prescription II that can be used to calculate radiative corrections to the inflationary potential owing to the inflaton couplings to bosons and fermions (section \ref{rad}). We also show that numerical results for coupling of the inflaton to bosons in prescription I (section \ref{press}) and couplings of the inflaton to bosons and fermions in prescription II (section \ref{pres}). Finally, we present our results and summary of them (section \ref{conc}).  
\section{Inflation with non-minimal coupling: Palatini formalism} \label{inf}

The Jordan frame Lagrangian density with non-minimally coupled scalar field $\phi$ with a canonical kinetic term and a potential $V_J(\phi)$:
\begin{equation} \label{vjphi}
\frac{\mathcal{L}_J}{\sqrt{-g}}=\frac12F(\phi)R-\frac12g^{\mu\nu}\partial_{\mu}\phi\partial_{\nu}\phi-V_J(\phi)\,,
\end{equation}
where the subscript $J$ indicates that the Lagrangian is defined in a Jordan frame and $F(\phi)=1+\xi \phi^2$. We are using units that the reduced Planck scale $m_P=1/\sqrt{8\pi G}\approx2.4\times10^{18}\text{ GeV}$ is set equal to unity, therefore we require $F(\phi)\to1$ after inflation.

In the metric formulation the connection is defined as a function of metric tensor, that is, Levi-Civita connection  ${\bar{\varGamma}={\bar{\varGamma}}(g^{\mu \nu})}$:
\begin{equation} \label{vargammametric}
\bar{\varGamma}_{\alpha \beta}^{\lambda}=\frac{1}{2}g^{\lambda \rho} (\partial_{a}g_{\beta \rho}+\partial_{\beta}g_{\rho \alpha}-\partial_{\rho}g_{\alpha \beta}).
\end{equation}
On the other hand, in the Palatini formalism both $g_{\mu \nu}$ and $\varGamma$ are independent variables, and the unique assumption is that the connection is torsion-free, $\varGamma_{\alpha \beta}^{\lambda}=\varGamma_{\beta \alpha }^{\lambda}$. If solving EoM, one can be obtained \cite{Bauer:2008zj}
\begin{eqnarray}\label{vargammapalatini}
\Gamma^{\lambda}_{\alpha \beta} = \overline{\Gamma}^{\lambda}_{\alpha \beta}
+ \delta^{\lambda}_{\alpha} \partial_{\beta} \omega(\phi) +
\delta^{\lambda}_{\beta} \partial_{\alpha} \omega(\phi) - g_{\alpha \beta} \partial^{\lambda} \omega(\phi),
\end{eqnarray}
where 
\begin{eqnarray}
\label{omega}
\omega\left(\phi\right)=\ln\sqrt{F(\phi)}.
\end{eqnarray}
Metric and Palatini formulation correspond to two different theories of gravity due to the difference of the connections in eqs. \eqref{vargammametric} and \eqref{vargammapalatini}. On the one hand, one more procedure of seeing differences in scalar field dynamics in Metric and Palatini approaches is to study the 	dilemma in the Einstein frame by virtue of the conformal transformation. 

To calculate the observational parameters, it is useful to switch to the Einstein (E) frame by applying a Weyl rescaling $g_{E, \mu \nu}=g_{J, \mu \nu}/F(\phi)$, so the Lagrangian density is obtained from \eqref{vjphi} \cite{Fujii:2003pa}:
\begin{equation} \label{LE}
\frac{\mathcal{L}_E}{\sqrt{-{g_E}}}=\frac12{R_E}-\frac{1}{2Z(\phi)}g_E^{\mu\nu}\partial_{\mu}\phi\partial_{\nu}\phi-V_E(\phi)\,,
\end{equation}
where
\begin{equation} \label{Zphi}
Z^{-1}(\phi)=\frac{1}{F(\phi)}\,,\qquad
V_E(\phi)=\frac{V_J(\phi)}{F(\phi)^2}\,,
\end{equation}
in the Palatini formulation. If we make a field redefinition
\begin{equation}\label{redefine}
\mathrm{d}\sigma=\frac{\mathrm{d}\phi}{\sqrt{Z(\phi)}}\,,
\end{equation}
we obtain the Lagrangian density for a
minimally coupled scalar field $\sigma$ with a canonical kinetic term. Thus, for the Palatini formulation, the field redefinition is induced just by the rescaling of the inflaton kinetic term indicated as eq. \eqref{redefine}. It does not depend on Jordan frame Ricci scalar. However, in Metric formulation, definiton of $\sigma$ depends on both the transformation of the Jordan frame Ricci scalar and the rescaling of the Jordan frame scalar field kinetic term \cite{Bauer:2008zj}. Therefore, the difference between two formulations corresponds to the different definition of $\sigma$ with the distinct non-minimal kinetic term including $\phi$.

For $F(\phi)=1+\xi\phi^2$, \eqref{Zphi} and \eqref{redefine} give this expressions:
\begin{description}
	\item[1.] Weak coupling limit \\ 
	If $|\xi|\phi^2\ll1$,  $\phi\approx\sigma$ and
	$V_J(\phi)\approx V_E(\sigma)$. Thus, the inflationary
	predictions are approximately the same as for minimal coupling in general. In that case, predictions are valid for both Metric and Palatini
	formulations.
	
	\item[2.] Large-field limit \\
	If $|\xi|\phi^2\gg1$, we have
	\begin{equation} \label{strong}
	\phi\simeq\frac{1}{\sqrt{\xi}}\sinh \left(\sigma\sqrt{\xi}\right),
	\end{equation}
	in the Palatini formulation. Eq. \eqref{strong} is different from Metric formulation result \cite{Bostan:2018evz,Bauer:2008zj}, in particular the presence of $\sqrt{\xi}$ in the argument of the hyperbolic function in the large-field limit. Using eq. \eqref{strong}, inflationary potential can be defined in terms of canonical scalar field $\sigma$, in accordingly one can be obtained slow-roll parameters in the Palatini formulation in large field limit according to $\sigma$.
    \end{description}

As long as the Einstein frame potential is obtained in terms of the canonical scalar field $\sigma$, observational parameters for inflation can be obtained using the slow-roll parameters \cite{Lyth:2009zz}
\begin{equation}\label{slowroll1}
\epsilon =\frac{1}{2}\left( \frac{V_{\sigma} }{V}\right) ^{2}\,, \quad
\eta = \frac{V_{\sigma \sigma} }{V}  \,, \quad
\xi ^{2} = \frac{V_{\sigma} V_{\sigma \sigma\sigma} }{V^{2}}\,,
\end{equation}
where $\sigma$'s in the subscript denote derivatives. Observational parameters such as the spectral index $n_s$, the tensor-to-scalar ratio $r$ and the running of the spectral index
$\mathrm{d} n_s/\mathrm{d} \ln k$ are given in the slow-roll
approximation by
\begin{equation}\label{nsralpha1}
n_s = 1 - 6 \epsilon + 2 \eta \,,\quad
r = 16 \epsilon, \quad
\alpha=\frac{\mathrm{d}n_s}{\mathrm{d}\ln k} = 16 \epsilon \eta - 24 \epsilon^2 - 2 \xi^2\,.
\end{equation}
In the slow-roll approximation, the number of e-folds is given by
\begin{equation} \label{efold1}
N_*=\int^{\sigma_*}_{\sigma_e}\frac{V\rm{d}\sigma}{V_{\sigma}}\,, \end{equation}
where the subscript ``$_*$'' denotes quantities when the scale
corresponding to $k_*$ exited the horizon, and $\sigma_e$ is the inflaton
value at the end of inflation, which we evaluate by $\epsilon(\sigma_e) =
1$. 

The amplitude of the curvature perturbation is given the form
\begin{equation} \label{perturb1}
\Delta_\mathcal{R}=\frac{1}{2\sqrt{3}\pi}\frac{V^{3/2}}{|V_{\sigma}|}.
\end{equation}
From the Planck measurement, best fit value for the pivot scale $k_* = 0.002$ Mpc$^{-1}$ is $\Delta_\mathcal{R}^2\approx   2.4\times10^{-9}$ \cite{Aghanim:2018eyx}.

In addition to this, we rewrite slow-roll parameters in terms of original field $\phi$ for numerical calculations. Using together with \eqref{redefine} and \eqref{slowroll1}, slow-roll parameters can be obtained in terms of $\phi$ \cite{Linde:2011nh}
\begin{equation}\label{slowroll2}  
\epsilon=Z\epsilon_{\phi}\,,\quad
\eta=Z\eta_{\phi}+{\rm sgn}(V')Z'\sqrt{\frac{\epsilon_{\phi}}{2}}\,,\quad
\xi^2=Z\left(Z\xi^2_{\phi}+3{\rm sgn}(V')Z'\eta_{\phi}\sqrt{\frac{\epsilon_{\phi}}{2}}+Z''\epsilon_{\phi}\right),
\end{equation}
where we defined 
\begin{equation}
\epsilon_{\phi} =\frac{1}{2}\left( \frac{V^{\prime} }{V}\right) ^{2}\,, \quad
\eta_{\phi} = \frac{V^{\prime \prime} }{V}  \,, \quad
\xi ^{2} _{\phi}= \frac{V^{\prime} V^{\prime \prime\prime} }{V^{2}}\,.
\end{equation}
Likewise, eqs. \eqref{efold1} and \eqref{perturb1} can be given by in terms of $\phi$
\begin{eqnarray}\label{perturb2}
N_*&=&\rm{sgn}(V')\int^{\phi_*}_{\phi_e}\frac{\mathrm{d}\phi}{Z(\phi)\sqrt{2\epsilon_{\phi}}}\,,\\
\label{efold2} \Delta_\mathcal{R}&=&\frac{1}{2\sqrt{3}\pi}\frac{V^{3/2}}{\sqrt{Z}|V^{\prime}|}\,.
\end{eqnarray}

To calculate the numerical values of observational parameters $n_s$, $r$ and $\alpha$, we  require a value of $N_*$ numerically. Assuming that a standard thermal history after inflation, $N_*$ is taken this form \cite{Liddle:2003as}
\begin{equation} \label{efolds}
N_*\approx64.7+\frac12\ln\frac{\rho_*}{m^4_P}-\frac{1}{3(1+\omega_r)}\ln\frac{\rho_e}{m^4_P}
+\left(\frac{1}{3(1+\omega_r)}-\frac14\right)\ln\frac{\rho_r}{m^4_P},
\end{equation}
where $\rho_{e}=(3/2)V(\phi_{e})$ is the
energy density at the end of inflation, $\rho_r$ is the energy density at the end of reheating and $\rho_*\approx V(\phi_*)$ is the energy density when the scale corresponding to $k_*$ exited the horizon. $\omega_r$ is the equation of state parameter during
reheating which is known as period to the oscillations of the inflaton. $\omega_r=1/3$ is a good approximation for the potentials which we take into account \cite{Bostan:2019fvk}, so $N_*$ is defined in that form 
\begin{equation} \label{efolds}
N_*\approx64.7+\frac12\ln\rho_*-\frac{1}{4}\ln\rho_e,
\end{equation}
does not depend on the reheat temperature. In section \ref{press} and \ref{pres}, we numerically calculate how the $n_s$ and $r$ values alter as a function of the coupling parameters $\kappa$ and $\xi$  by taking into consideration two different prescription. We also show numerically that the change in running of the spectral index as a function of $\kappa$ for selected $\xi$ values. The calculation method is in such a way: We form a grid of points in the $\kappa-\xi$ plane. For every ($\kappa$, $\xi$) point, we begin the compute with attribute to an initial self-coupling constant $\lambda$ value. We then determine numerical values of $\phi_{e}$ using $\epsilon(\phi_e)=1$, and we obtain $\phi_*$ using eq. \eqref{efold2}. The e-fold number $N_*$ is computed using eq. \eqref{perturb2} and compared with eq. \eqref{efolds}. The initial value of self-coupling constant $\lambda$ is then set and the calculation is replicated until the two $N*$ values become equal. The $\phi_*$ value calculated this way is inserted in eqs. \eqref{slowroll2} and \eqref{nsralpha1} to obtain the $n_s$, $r$ and $\mathrm{d} n_s/\mathrm{d} \ln k$ values. Finally, the calculation is replicated over the entire grid, with self-coupling constant $\lambda$ solutions for every point used as initial values of their neighbours.

\section{Overview of radiative corrections} \label{rad}
Couplings of the inflaton with other fields lead to radiative corrections in the inflationary potential. These corrections can be expressed by CW one-loop corrections which define the interaction of inflaton at the leading order, with the quantum fields, $\chi$ is the scalar boson and $\Psi$ is the Dirac fermion. CW one-loop potential is defined as \cite{Coleman:1973jx,Enqvist:2013eua,Weinberg:1973am}
\begin{eqnarray} \label{CW}
\Delta V(\phi)=\sum\limits_{i}\frac{(-1)^F}{64\pi^2}M_i(\phi)^4 \ln\Big(\frac{M_i(\phi)^2}{\mu^2}\Big).
\end{eqnarray}
Here, $F$ takes for bosons (fermions) $+1$ $(-1)$ values. $\mu$ is a renormalization scale and $M_i(\phi)$ indicates field dependent mass. 

To begin with, we consider the potential terms for a minimally coupled quartic potential interacts to other scalar $\chi$ and to a Dirac fermion $\Psi$:
\begin{eqnarray}\label{lag}
V(\phi,\chi,\Psi)= \frac{\lambda}{4}\phi^4+h\phi\bar{\Psi}\Psi+m_\Psi\bar{\Psi}\Psi+\frac{1}{2}g^2\phi^2\chi^2+\frac{1}{2}m_\chi^2\chi^2 .
\end{eqnarray}
Under the assumptions
\begin{eqnarray} \label{assume}
g^2\phi^2\gg m_\chi^2, \qquad g^2\gg\lambda, \qquad h\phi\gg m_\Psi, \qquad h^2\gg\lambda,
\end{eqnarray} 
the inflationary potential which includes of the  CW one-loop corrections given by eq. \eqref{CW} can be obtained in that form 
\begin{eqnarray}\label{coupling}
V(\phi)\simeq\frac{\lambda}{4}\phi^4\pm\kappa\phi^4\ln\left(\frac{\phi}{\mu}\right),
\end{eqnarray} 
where the $ + $ $(-)$ sign corresponds to the bosons
(fermions) case where the coupling is dominant and we have described the radiative correction coupling parameter
\begin{eqnarray}\label{kappa}
\kappa\equiv\frac{1}{32\pi^2}\Big|(g^4-4h^4)\Big|.
\end{eqnarray}
Note that the potential in eq. \eqref{coupling} is approximation which can be obtained from the one-loop renormalization group improved effective actions, see for instance ref. \cite{Okada:2010jf}. What is more, the one-loop term can be larger than the tree-level term while two-loop corrections. This is because the one-loop term comes from a distinct interaction and not from the self-interaction of the inflaton field \cite{Weinberg:1973am}. Furthermore, generalizing eq. \eqref{coupling} to the non-minimally coupled case is confronted with an uncertainty if the UV completion of the low-energy effective theory is not defined clearly, as debated in ref. \cite{Bezrukov:2013fka}.

In the literature, two different prescriptions typically for the computation of radiative corrections are considered. Two prescriptions are called as prescription I \cite{George:2013iia,Bezrukov:2013fka} and prescription II \cite{DeSimone:2008ei,Barvinsky:2009fy,Barvinsky:2009ii}. The difference between these two renormalization prescriptions is that the field dependent masses are defined differently. In summary: 
\begin{description}
	\item[1.]  In prescription I, the field-dependent masses expressed in the Einstein frame. 
	\item[2.]  In prescription II, the field-dependent masses expressed in the Jordan frame. 
\end{description}	  
In prescription I, the field dependent masses in the
one-loop CW potential are defined in the Einstein frame. Using the transformations
\begin{eqnarray}
V(\phi)=\frac{V_J(\phi)}{F(\phi)^2}, \ \ \tilde{\phi}=\frac{\phi}{\sqrt{F(\phi)}}, \ \ \tilde{\Psi}=\frac{\Psi}{F(\phi)^{3/4}},\ \ \tilde{m}_{\Psi}(\phi)=\frac{m_\Psi(\phi)}{\sqrt{F(\phi)}}, \ \ \tilde{m}^2_{\chi}=\frac{m^2_\chi}{F(\phi)}.
\end{eqnarray} 
Thus, in prescription I, the one-loop corrected potential can be obtained in the Einstein frame in that form
\begin{eqnarray}\label{p1}
V(\phi)=\frac{\frac{\lambda}{4}\phi^4\pm\kappa \phi^4 \ln\Big(\frac{\phi}{\mu \sqrt{1+\xi \phi^2}}\Big)}{(1+\xi\phi^2)^2} .
\end{eqnarray}
In prescription II, the field dependent masses in the one-loop CW potential are defined in the Jordan frame, thus eq. \eqref{coupling} corresponds to the one-loop CW potential in the Jordan frame. As a result, for prescription II, the Einstein frame potential in this case is given by
\begin{eqnarray} \label{p2}
V(\phi)=\frac{\frac{\lambda}{4}\phi^4\pm\kappa \phi^4 \ln\Big(\frac{\phi}{\mu}\Big)}{(1+\xi\phi^2)^2} .
\end{eqnarray} 
It should be emphasized that changing the value of $\mu$ does not vary for the potential form of eqs. \eqref{p1} and \eqref{p2}. Forms of these potentials change solely with the shift of $\lambda$. Therefore $n_s$, $r$ and $\alpha$ values do not change with the value of $\mu$ for fixed values of $\kappa$ and $\xi$.

Finally, note that here, the potentials in eqs. \eqref{p1} and \eqref{p2} are approximations which can be acquired from the one-loop renormalization group improved effective actions, see e.g. ref \cite{Okada:2010jf}.  
\section{Quartic potential with CW one-loop corrections in the Palatini formalism: Prescription I}
 \label{press}
 \begin{figure}[t!]
 	\centering
 	\includegraphics[angle=0, width=10cm]{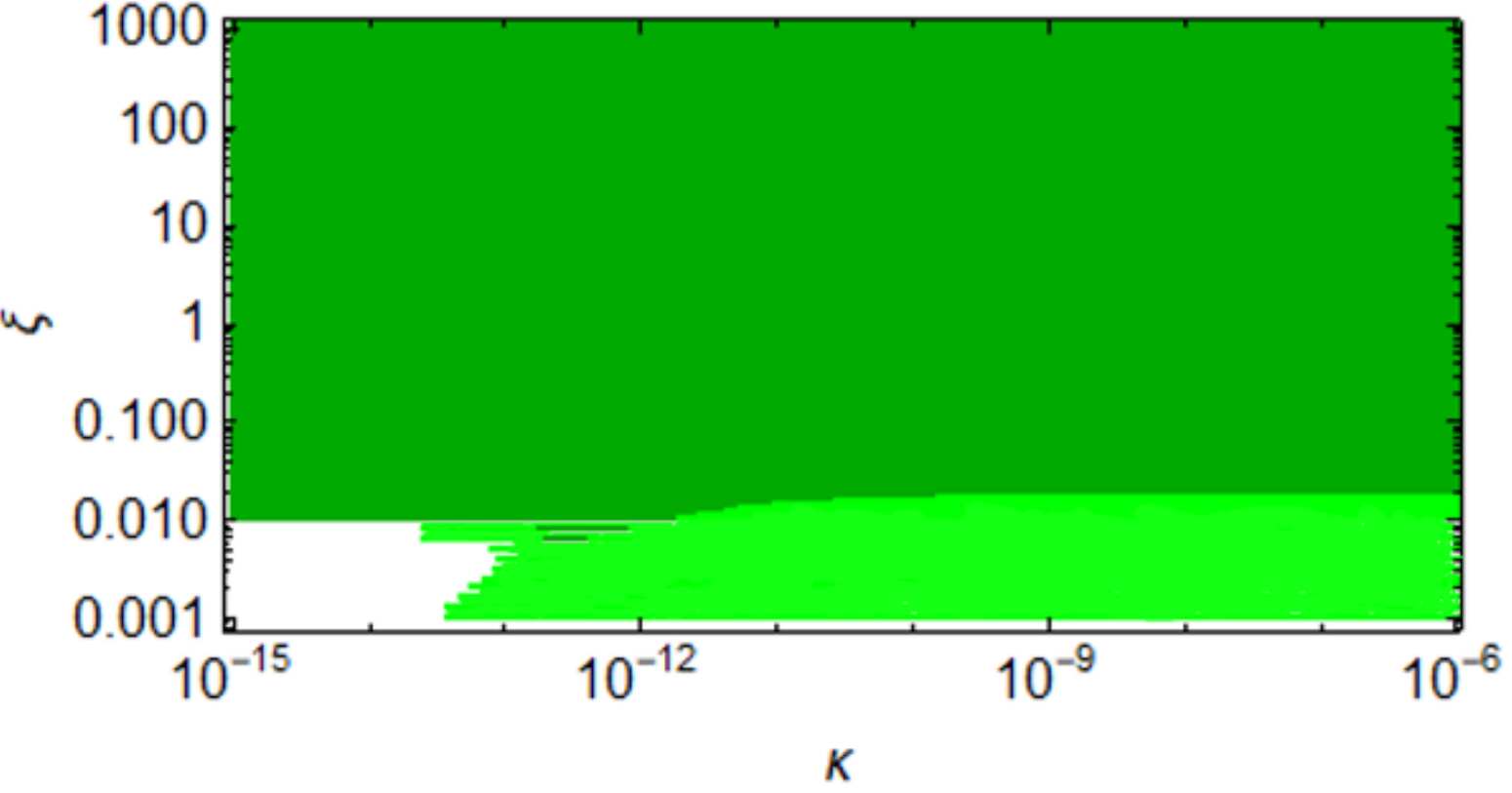}
 	\includegraphics[angle=0, width=17cm]{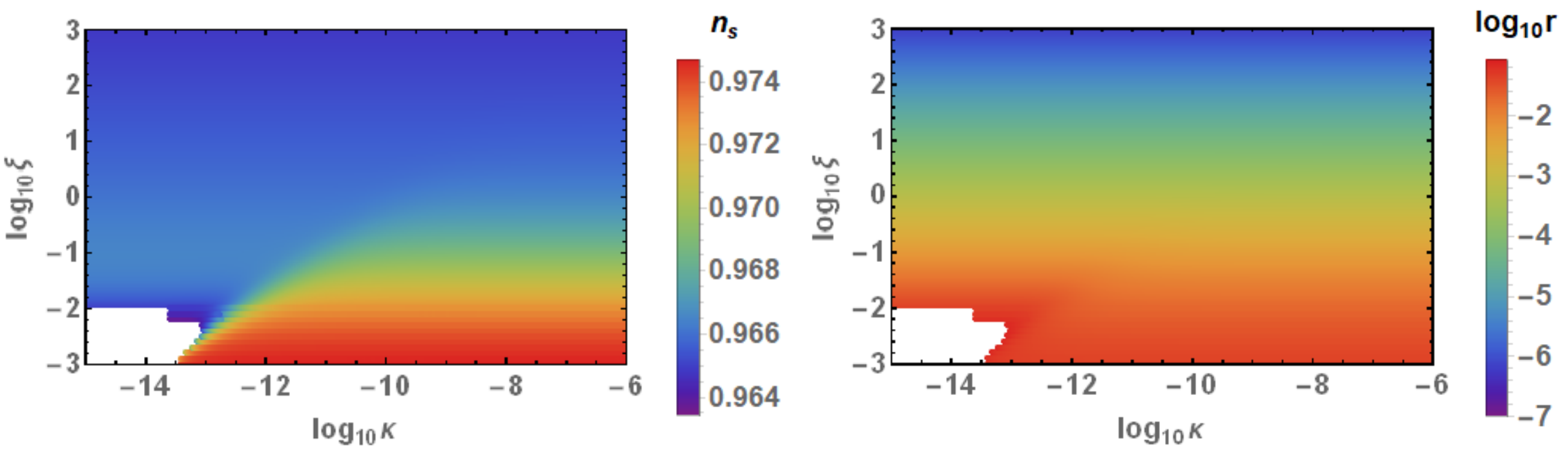}
 	\caption{ For prescription I and inflaton coupling to bosons in Palatini formalism, the top figure displays in light green
 		(green) the regions in the $\kappa-\xi$ plane predict $n_s$ and $r$ values inside the 95\% (68\%) CL contours
 		based on data taken by the Keck Array/BICEP2 and Planck collaborations \cite{Ade:2018gkx}. Bottom figures display
 		$n_s$ and $r$ values in these regions.
 	}
 	\label{fig1}
 \end{figure}
\begin{figure}[]
	\centering
	\includegraphics[angle=0, width=13cm]{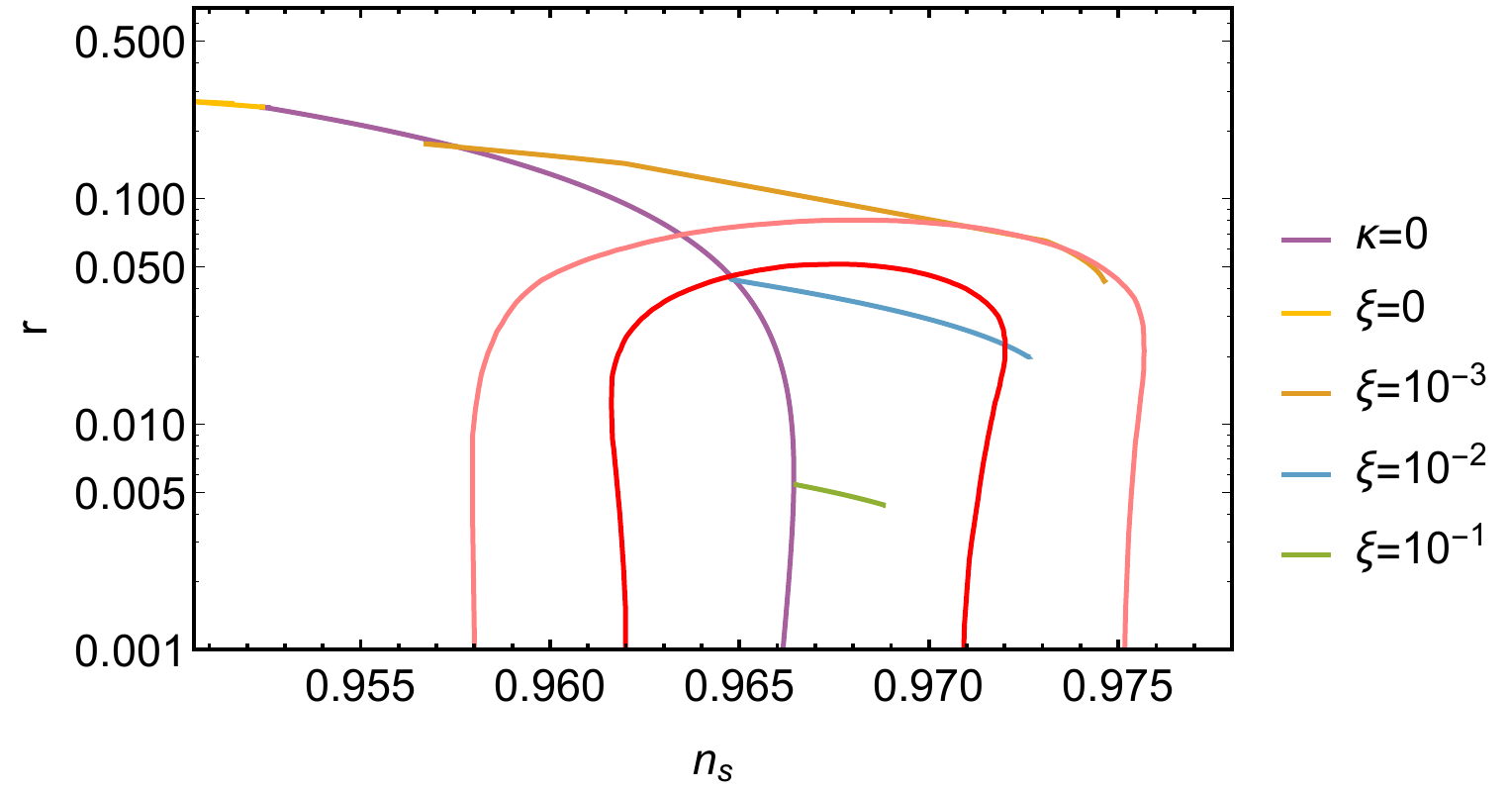}
	\includegraphics[angle=0, width=15cm]{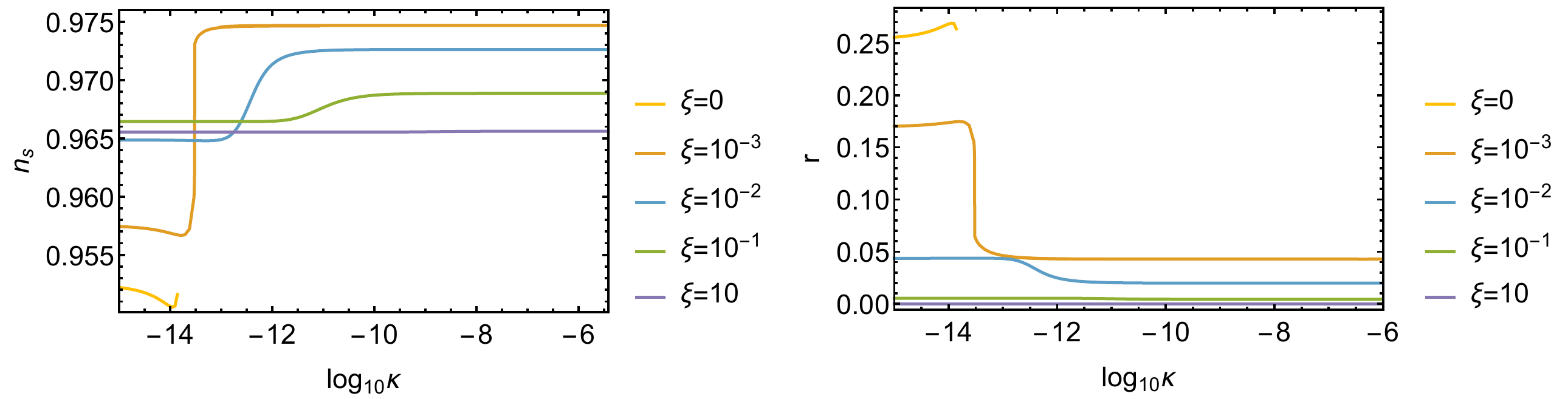}
	\includegraphics[angle=0, width=10cm]{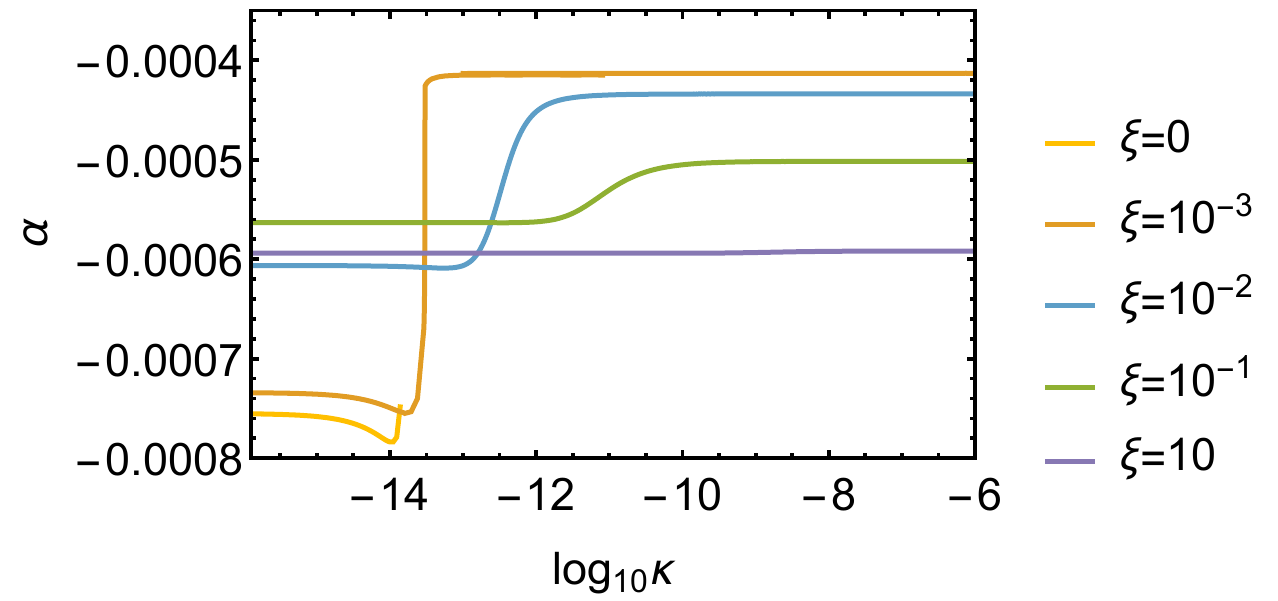}
	\caption{ For prescription I and inflaton coupling to bosons in Palatini formalism, the change in $n_s$, $r$ and $\alpha$ as a function
		of $\kappa$ is plotted for selected $\xi$ values. The pink (red) contour in the top figure corresponds to the 95\% (68\%) CL contour based on data taken by the Keck Array/BICEP2 and Planck collaborations \cite{Ade:2018gkx}.
	}
	\label{fig2}
\end{figure}	
In this section we numerically calculate how the $n_s$ and $r$ values change as a function of the coupling parameters $\xi$ and $\kappa$ by taking into account to the potential in eq. \eqref{p1}, with a $+$ sign for the inflaton dominantly coupling to bosons in Palatini formulation. We also demonstrate predict the running of spectral index for selected $\xi$ values for the inflaton coupling to bosons in prescription I. For prescription I and inflaton coupling to bosons, figure \ref{fig1} demonstrates the region in the $\xi$ and $\kappa$ plane where $n_s$, $r$ values are agreement with the current data. Figure \ref{fig2} shows that how $n_s$, $r$ and $\alpha$ values shift with coupling parameter $\kappa$ for selected $\xi$ values. It can be seen from the figures that $n_s$ and $r$ values depend on more sensitively to the value of $\xi$ instead of $\kappa$. Similar to the results of Metric formulation \cite{Bostan:2019fvk}, as $\kappa$ is increased taking $\xi$ fixed, there is a shifting in $n_s$ and $r$ values for a comparatively narrow range of $\kappa$. $n_s$ and $r$ no longer vary at even bigger $\kappa$ values, on the one hand latest result is confronted with some caveats as debated below. Furthermore, we can see that from figure \ref{fig2}, \eqref{perturb2}, \eqref{efold2} and \eqref{efolds} can be supplied simultaneously for arbitrarily large values of coupling parameter $\kappa$, this case is the same as Metric formulation results again. On the other hand, as we mentioned at the end of section \ref{rad}, the potential that we take into account is an approximation for the one-loop renormalization group improved effective action and this approximation will break down for larger values of $\kappa$ at last. What is more, higher loop corrections will become significant eventually. For large values of coupling parameter $\kappa$, inflationary solutions can be acquired for fine tuned values of the coupling parameters solely. To explain this case, we analyze the potential in eq. \eqref{p1} in $\xi \phi^2\gg1$ limit and take $\mu=1$ for simplicity. In that approximation, the potential in eq. \eqref{p1} can be obtained that form
\begin{eqnarray} \label{quartic}
V(\phi)=\frac{ A \phi^4}{(1+\xi \phi^2)^2},
\end{eqnarray}
where $A\equiv\lambda/4-(\kappa/2)\ln \xi$. Furthermore, the potential in eq. \eqref{quartic} approaches to the well-known potential in the literature which is non-minimal quartic inflation potential on condition that with $(\lambda/4)$ switched to $A$ in eq. \eqref{quartic}. Using eq. \eqref{strong}, this potential can be computed in the large-field limit $\xi \phi^2\gg1$ in terms of $\sigma$ in that form 
\begin{eqnarray} \label{p1sigma}
V(\sigma)\approx \frac{A}{\xi^2} \left[1-8 \exp\left(-2\sqrt{\xi}\sigma \right)\right].
\end{eqnarray}
Using eq. \eqref{efold1}, $\exp(2\sqrt{\xi}\sigma )\approx 32\xi N$. To conclude, using eq. \eqref{perturb1}, we find that 
\begin{eqnarray} \label{p1lambda}
\lambda \approx \frac{12 \pi^2  \Delta^2_R \xi  }{ N^2}+ 2\kappa \ln \xi.
\end{eqnarray}
The first term in the right side of the eq. \eqref{p1lambda} is virtually $8\times 10^{-11}\xi$ in the case of large-field limit. On the condition that $2\kappa \ln\xi$ is bigger than this term, eq. \eqref{p1lambda} can only be satisfied if $\lambda$ almost fairly equals  $2\kappa \ln\xi$. 
Furthermore, as it can be seen that from figures \ref{fig1} and \ref{fig2}, $r$ takes very small values for $\xi\gg 1$ cases which is the pivotal difference between Metric and Palatini formulation for inflaton to bosons coupling in prescription I,  to examine the results in that coupling type in prescription I for Metric formulation see ref. \cite{Bostan:2019fvk}. We can say that $r$ is extremely suppressed in the Palatini formulation for large $\xi$ values for the inflaton to bosons coupling in prescription I similar to the results of other refs. \cite{Bauer:2008zj,Rasanen:2017ivk,Racioppi:2017spw}. 

Finally, we also display the running of spectral index for inflaton to bosons coupling in prescription I in figure \ref{fig2} as a function of $\kappa$ for selected $\xi$ values. It can be seen that $\alpha$ values are too tiny to be detected in the future experimental for all chosen $\xi$ values. 
\section{Quartic potential with CW one-loop corrections in the Palatini formalism: Prescription II} \label{pres}

In this section, we illustrate numerically how the $n_s$ and $r$ values vary as a function of the coupling parameters $\kappa$ and $\xi$ in Palatini formalism, by taking into account to the potential in eq. \eqref{p2}, with a $+$ $ (-)$ sign for the inflaton dominantly coupling to bosons (fermions). Similarly, in prescription I, we show that predict the running of spectral index for selected $\xi$ values for the inflaton coupling to bosons and also fermions in prescription II.
\begin{figure}[]
	\centering
	\includegraphics[angle=0, width=12cm]{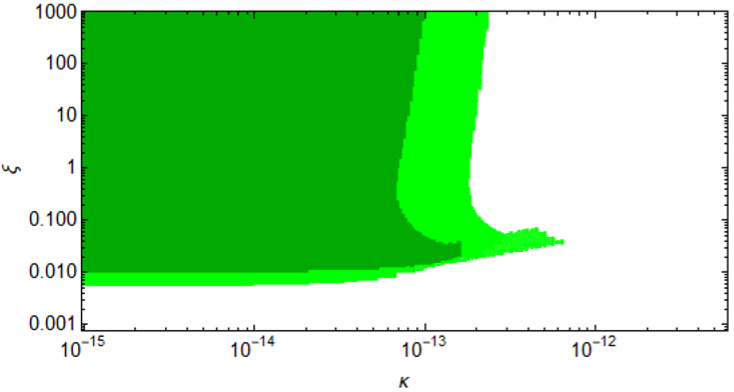}
	\includegraphics[angle=0, width=16cm]{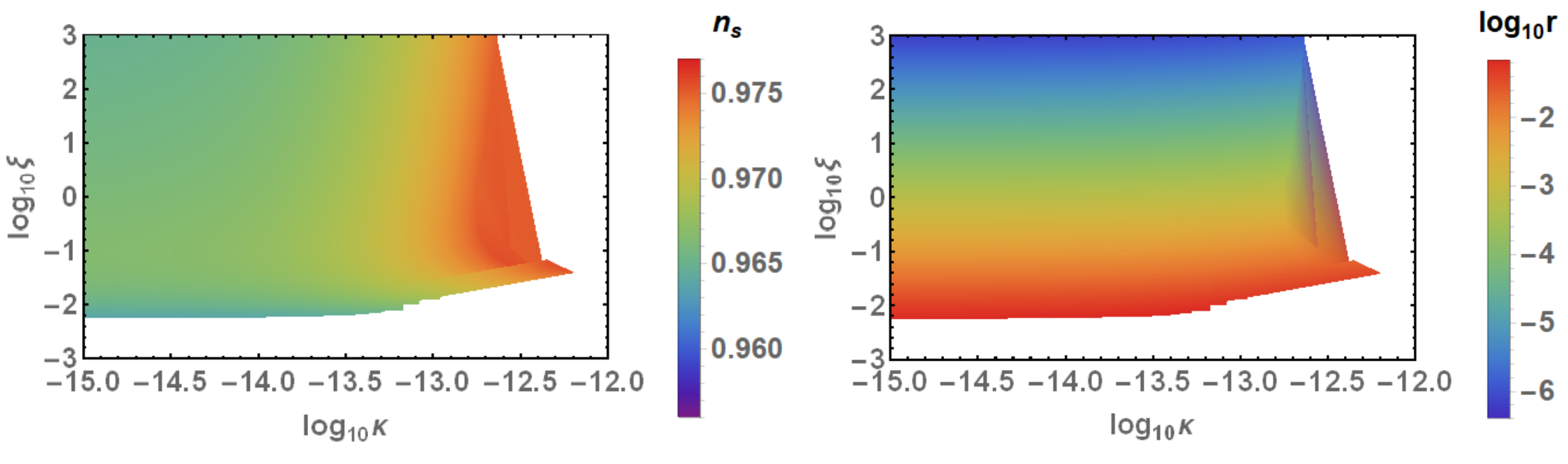}
	\caption{ For prescription II and inflaton coupling to bosons in Palatini formalism, the top figure displays in light green
		(green) the regions in the $\kappa-\xi$ plane predict $n_s$ and $r$ values inside the 95\% (68\%) CL contours
		based on data taken by the Keck Array/BICEP2 and Planck collaborations \cite{Ade:2018gkx}. Bottom figures display
		$n_s$ and $r$ values in these regions.
	}
	\label{fig3}
\end{figure}
For prescription II and inflaton coupling to bosons, figure \ref{fig3} exhibits the region in the $\kappa$ and $\xi$ plane where $n_s$ and $r$ values are agreement with the current data. Figure \ref{fig4} displays that how $n_s$, $r$ and $\alpha$ values change with $\kappa$ for chosen $\xi$ values. Unlike to the inflaton to bosons coupling in prescription I discussed in previous section, inflaton to bosons coupling in prescription II has a $\kappa_{\mathrm{max}}$, that is, the maximum $\kappa$ values that permit a solution of eqs. \eqref{perturb2}, \eqref{efold2} and \eqref{efolds} simultaneously which are also demonstrated in figure \ref{fig4} for selected $\xi$ values. From the figures also it can be seen that for $10^{-2}\lesssim\xi\lesssim10^{-1}$, the $n_s$ and $r$ values saturating the linear limit as $\kappa$ approaches $\kappa_{\mathrm{max}}$. This saturating to the linear limit for $\xi\gtrsim 10^{-1}$ values to the CW inflation in Palatini formalism emphasized in ref. \cite{Racioppi:2017spw}.  On the other hand, it is clear from the figures that for $\xi\gg1$ values, linear inflation predictions are lost because $r$ values are very tiny and therefore in Palatini formulation, $r$ is much smaller than the Metric formulation for the inflaton coupling to bosons for $\xi\gg1$, the inflaton coupling to bosons results in Metric formulation was shown in ref. \cite{Bostan:2019fvk}. Moreover, for $\kappa>\kappa_{\mathrm{max}}$, there is no solution, because eqs. \eqref{perturb2}, \eqref{efold2} and \eqref{efolds} cannot be satisfied simultaneously. Finally, for prescription II and inflaton coupling to bosons, predict a running of the spectral index that is too tiny to be observed in the near future experimental, as it can be seen from the figure \ref{fig4}. 

\begin{figure}[t!]
	\centering
	\includegraphics[angle=0, width=14cm]{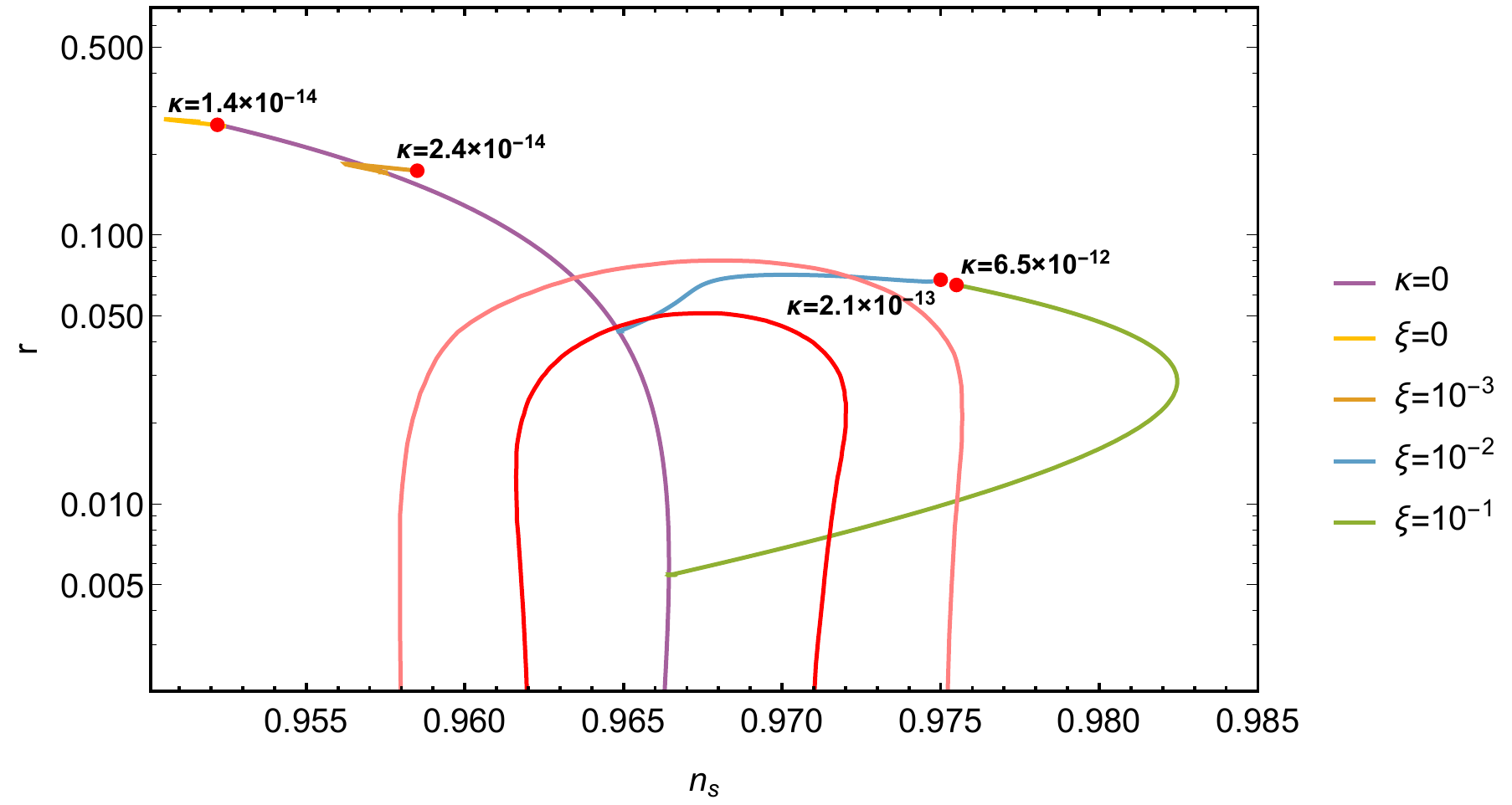}
	\includegraphics[angle=0, width=16cm]{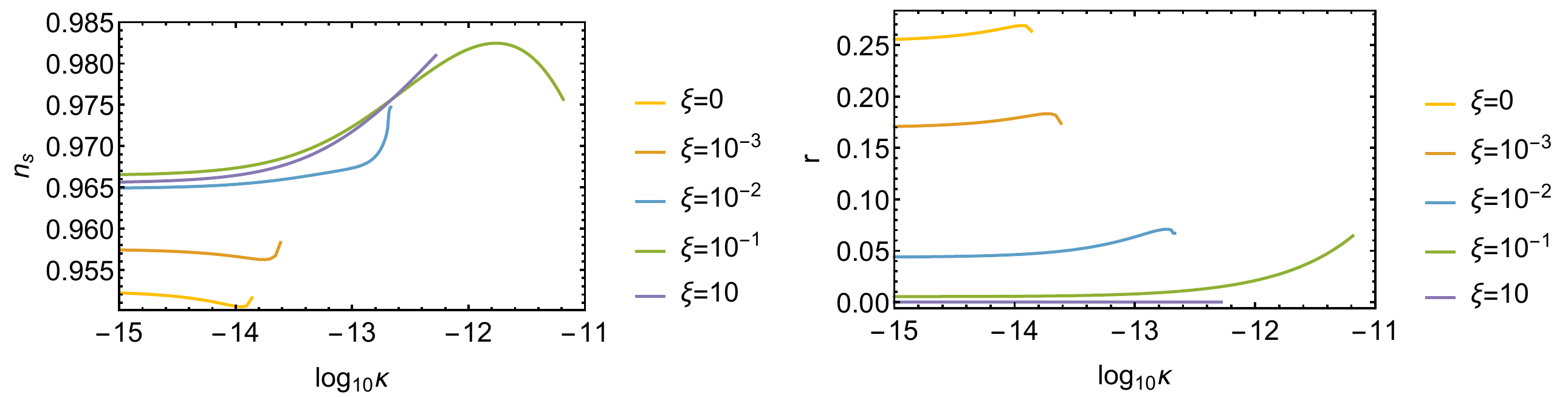}
	\includegraphics[angle=0, width=11cm]{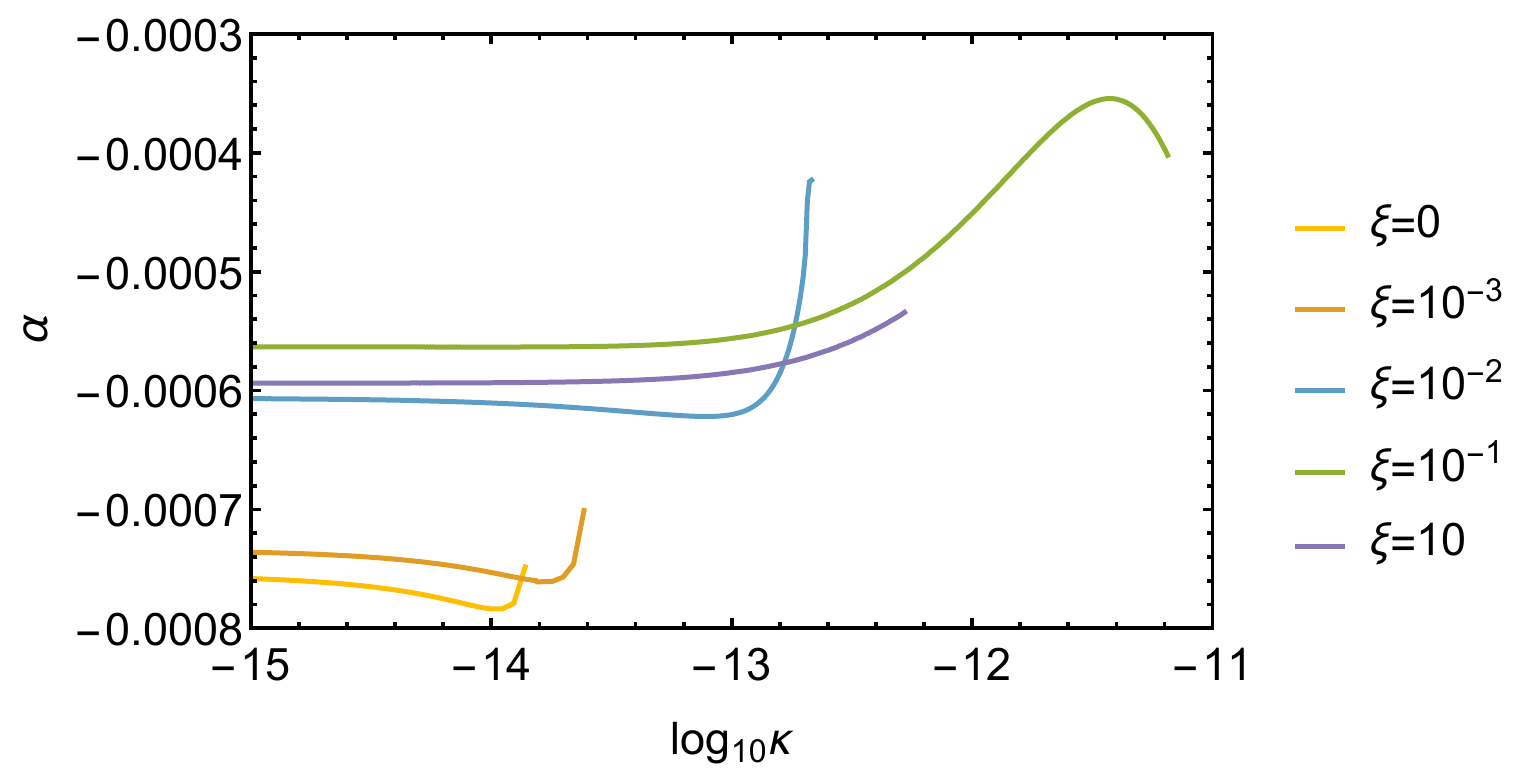}
	\caption{For prescription II and inflaton coupling to bosons in Palatini formalism, the change in $n_s$, $r$ and $\alpha$ as a function
		of $\kappa$ is plotted for selected $\xi$ values. The pink (red) contour in the top figure corresponds to the 95\% (68\%) CL contour based on data taken by the Keck Array/BICEP2 and Planck collaborations \cite{Ade:2018gkx}.
		The red points display the maximum $\kappa$ values. These values, increasing with $\xi$, are also noted in the
		figure.
}
	\label{fig4}
\end{figure}
\begin{figure}[t!]
	\centering
	\includegraphics[angle=0, width=10cm]{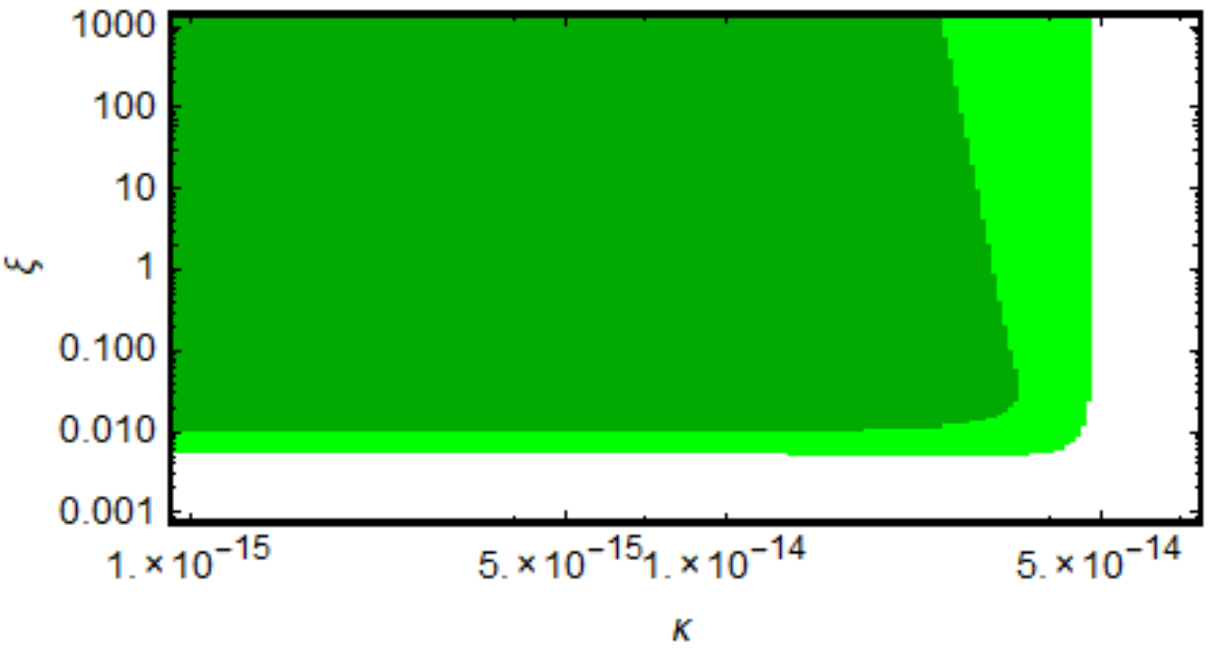}
	\includegraphics[angle=0, width=16cm]{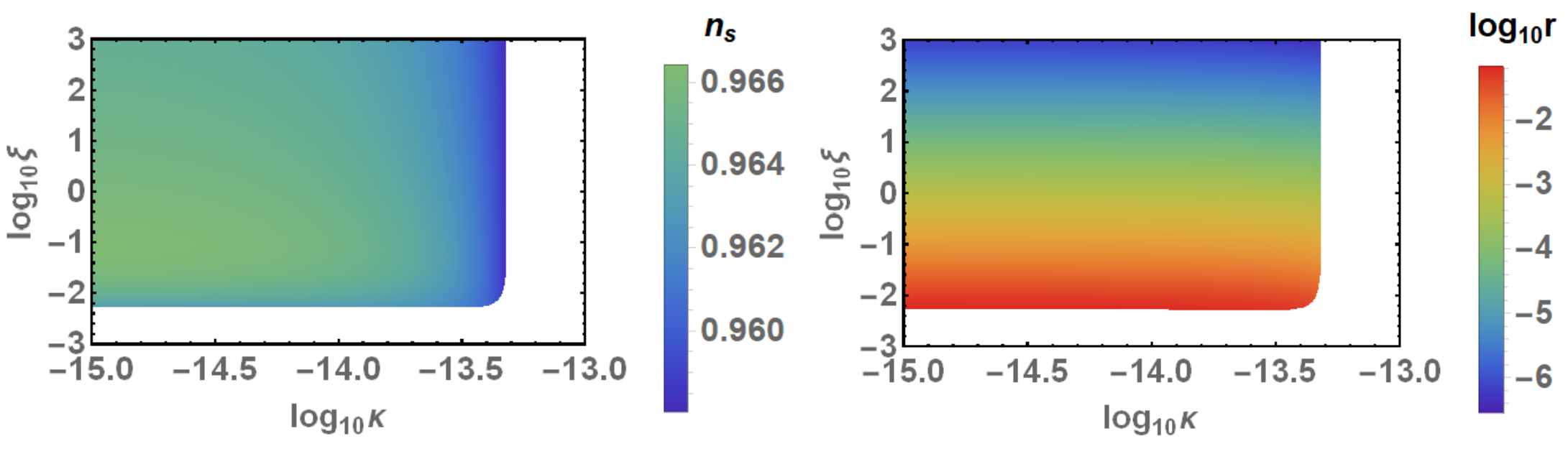}
	\caption{ For prescription II and inflaton coupling to fermions in Palatini formalism, the top figure displays in light green
		(green) the regions in the $\kappa-\xi$ plane predict $n_s$ and $r$ values are inside the 95\% (68\%) CL contours
		based on data taken by the Keck Array/BICEP2 and Planck collaborations \cite{Ade:2018gkx}. Bottom figures display
		$n_s$ and $r$ values in these regions.
	}
	\label{fig5}
\end{figure}

\begin{figure}[t!]
	\centering
	\includegraphics[angle=0, width=12cm]{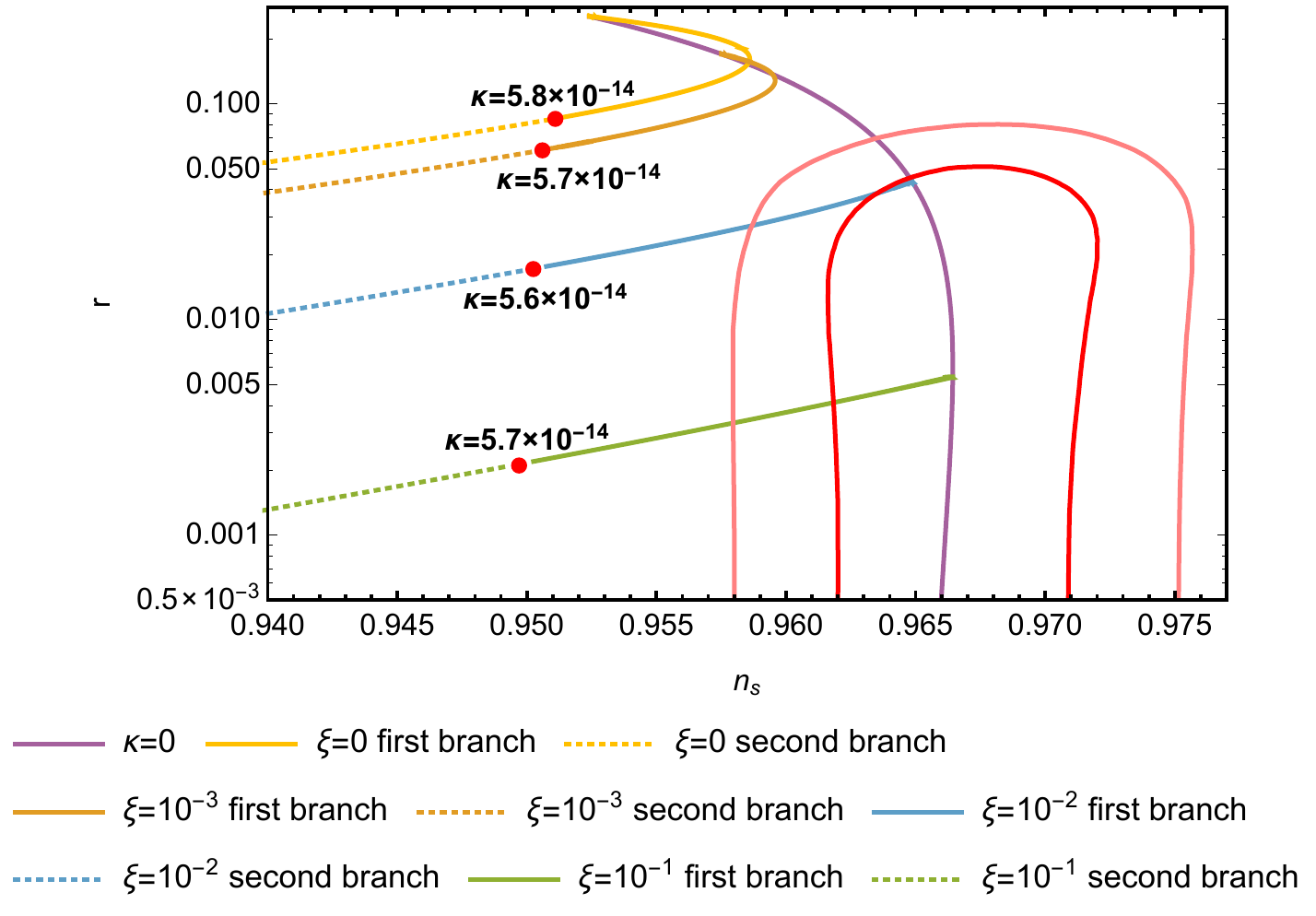}
	\includegraphics[angle=0, width=16cm]{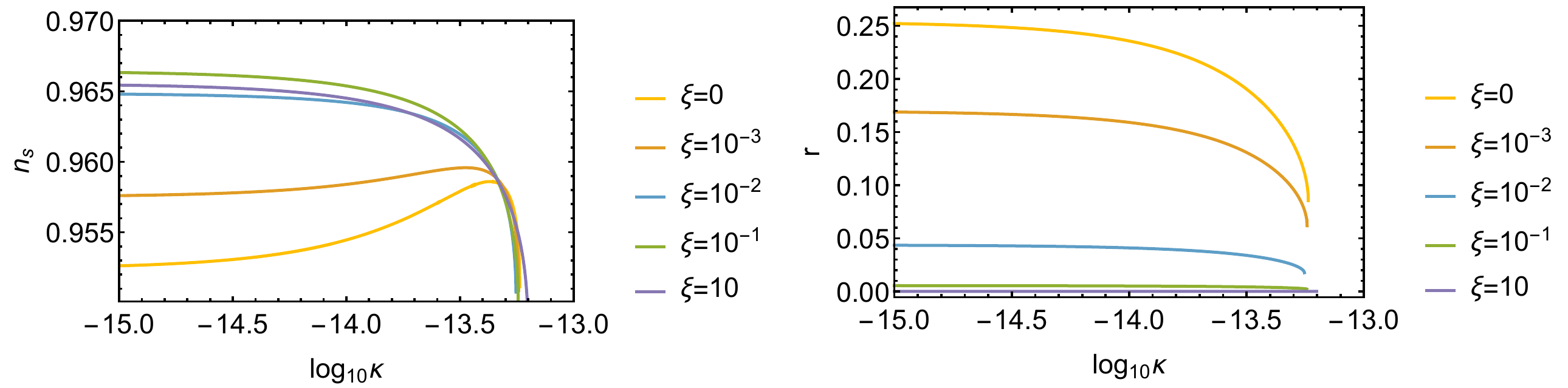}
	\includegraphics[angle=0, width=10cm]{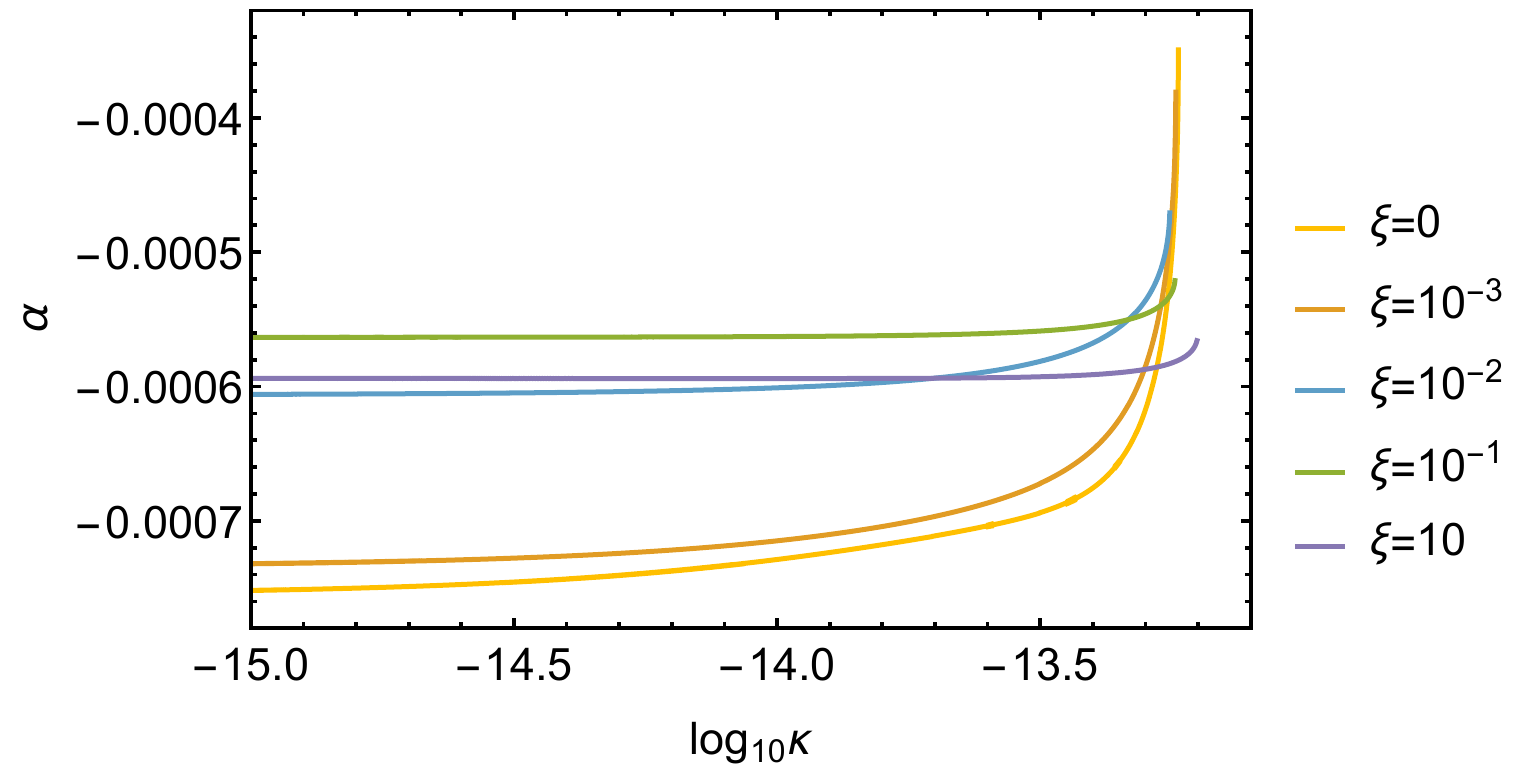}
	\caption{For prescription II and inflaton coupling to fermions, the change in $n_s$, $r$ and $\alpha$ as a function of $\kappa$ is plotted for selected $\xi$ values in Palatini formalism. The pink (red) contour in the top figure corresponds to the 95\% (68\%) CL contour based on data taken by the Keck Array/BICEP2 and Planck collaborations \cite{Ade:2018gkx}. The solid (dotted) portions of the curves correspond to first (second) branch of solutions. The red points display the maximum $\kappa$ values where the two branch of solutions meet. These values are also noted in the figure. The bottom figures only display the first branch solutions.
}
	\label{fig6}
\end{figure}

	
For prescription II and inflaton coupling to fermions, figure \ref{fig5} shows that the region in the $\kappa$ and $\xi$ plane where $n_s$ and $r$ values are agreement with the current data for solutions of the first branch. Also it can be seen that from figure \ref{fig5}, noticeable change occurs in the $n_s$ and $r$ values, when $\kappa$ values become the same order of magnitude as $\kappa_{\mathrm{max}}$. The inflaton coupling to fermions case investigated in ref. \cite{NeferSenoguz:2008nn} taking $\xi=0$ in Metric formulation. There it was demonstrated that there are two solutions for each $\kappa$ value that is lower than a $\kappa_{\mathrm{max}}$ value. As it can be seen from figure \ref{fig6} in our work, there are two solutions also $\xi=0$ and $\xi\neq0$ for our selected $\xi$ values in Palatini formalism. We indicate the branch of solutions with larger $\lambda$ for an endowed $\kappa$ as the first branch, and another branch of solutions as the second branch. In that case, there are two branch of solutions but the second branch of solutions are not agreement with the current data at value of neither $\xi$ nor $\kappa$. The first branch solutions replace from the red points towards the $\kappa=0$ curve as $\kappa$ declines. On the contrary, the second branch solutions shift towards small $n_s$ values. For prescription II and inflaton coupling to fermions investigated in Metric formulation refs. \cite{Okada:2010jf,Bostan:2019fvk}, as we mentioned before. We can say that our results for the inflaton coupling to fermions in prescription II agree with refs. \cite{Okada:2010jf,Bostan:2019fvk} except for $\xi\gg1$ cases since $r$ is very smaller for values of $\xi\gg1$ in Palatini approach results than the Metric one. 

Similarly to the other cases, predict a running of the spectral index that is too tiny to be observed in the near future, it can be seen from figure \ref{fig6}. 
\clearpage	
\section{Conclusion}\label{conc}
In this paper, we studied the inflationary predictions with non-minimal coupling in Palatini formulation in section \ref{inf} and then we shortly discussed the radiative corrections to the potential for two different renormalization prescriptions in section \ref{rad}. We then displayed numerically the effect of radiative corrections on the $n_s$, $r$ and $\alpha$ in Palatini formalism to inflaton coupling to bosons for prescription I in section \ref{press} and inflaton coupling to bosons and fermions for prescription II in section \ref{pres}. 

In general, we presented that while the radiative corrections 	restrain inflation with an adequate duration after a $\xi$ depending upon maximum value $\kappa_{\mathrm{max}}$ of the coupling parameter $\kappa$ specified in eq. \eqref{kappa} and also radiative corrections do not alter $n_s$ and $r$ values importantly if $\kappa$ is not the same order of magnitude as $\kappa_{\mathrm{max}}$. On the other hand, differently from the other cases, for the inflaton coupling to bosons in prescription I \eqref{perturb2}, \eqref{efold2} and \eqref{efolds} eqs. can be satisfied simultaneously for large values of $\kappa$ but as clarified in section \ref{rad}, we took into account this result as a structure of the approximation we analyzed for the potential. 

We also investigated for $\xi\gg1$ values in Palatini approach, $r$ has very tiny value for both inflaton to bosons coupling in prescription I and inflaton to bosons and fermions coupling in prescription II. In these cases, $r$ also has very smaller values for $\xi\gg1$ than the Metric formulation, as explained in section \ref{press} and section \ref{pres}. We observed for values of $10^{-2}\lesssim\xi\lesssim10^{-1}$, the $n_s$ and $r$ values saturating the linear limit as $\kappa$ approaches $\kappa_{\mathrm{max}}$ to the inflaton coupling to bosons for prescription II in section \ref{pres}. What is more, we demonstrated all the cases considered in Metric formulation the prominent Starobinsky attractor is lost in Palatini formulation. We consider that for $\xi\gg1$ cases, the difference between Metric formulation and Palatini formulation can be distinguished by the precision measurements \cite{Remazeilles:2017szm} of future experimental. However, on the assumption that a larger amount of $r$ is obtained, Palatini approach can eliminate naturally. 

Finally, we found that in section \ref{press} and section \ref{pres} for Palatini formulation the predict a running of the spectral index $\mathrm{d} n_s/\mathrm{d} \ln k$ that is too tiny to be observed in the near future experimental for both inflaton coupling to bosons in prescription I and inflaton couplings to bosons and fermions in prescription II for selected $\xi$ values. 
\acknowledgments

The author very thanks V. Nefer \c{S}eno\u{g}uz.


\end{document}